\documentclass{aastex62}

\received{August 7, 2018}
\revised{September 6, 2018}
\accepted{September 18, 2018}
\submitjournal{ApJ}

\shorttitle{Unravelling VVV-WIT-06}
\shortauthors{Banerjee et al.}

\begin{document}

\title{Unravelling the infrared transient VVV-WIT-06: the case for an origin in a classical nova\footnote{Released on August, 7th, 2018}}

\correspondingauthor{D.P.K. Banerjee}
\email{orion@prl.res.in}

\author{D.P.K. Banerjee}
\affil{Astronomy and Astrophysics Division, Physical Research Laboratory, Navrangpura, Ahmedabad, India, 380009}
\author{E. Y. Hsiao}
\affiliation{Department of Physics, Florida State University, Tallahassee, FL 32306, USA}
\author{T. Diamond}
\affiliation{NASA/GSFC, Mail Code: 665, Greenbelt , MD 20771, USA}
\author[0000-0002-1296-6887]{L. Galbany}
\affiliation{PITT PACC, Department of Physics and Astronomy, University of Pittsburgh, Pittsburgh, PA 15260, USA}
\author{N. Morrell}
\affiliation{Las Campanas Observatory, Carnegie Observatories, Casilla 601, La Serena, Chile}
\author{D. Minniti}
\affiliation{Departamento de Ciencias F\'isicas, Facultad de Ciencias Exactas, Universidad Andr\'es Bello, Av. Fernandez Concha 700, Las Condes, Santiago, Chile}
\affiliation{Millennium Institute of Astrophysics, Av. Vicuna Mackenna 4860, 782-0436, Santiago, ChilE}
\affiliation{Vatican Observatory, V-00120 Vatican City State, Italy}
\author{H. Kuncarayakti}
\affiliation{Finnish Centre for Astronomy with ESO (FINCA), University of Turku, V\"{a}is\"{a}l\"{a}ntie 20, 21500 Piikki\"{o}, Finland}
\affiliation{Tuorla Observatory, Department of Physics and Astronomy, University of Turku, V\"{a}is\"{a}l\"{a}ntie 20, 21500 Piikki\"{o}, Finland}
\author{S. Mattila}
\affiliation{Tuorla Observatory, Department of Physics and Astronomy, University of Turku, V\"{a}is\"{a}l\"{a}ntie 20, 21500 Piikki\"{o}, Finland}
\author{J. Harmanen}
\affiliation{Tuorla Observatory, Department of Physics and Astronomy, University of Turku, V\"{a}is\"{a}l\"{a}ntie 20, 21500 Piikki\"{o}, Finland}

\begin{abstract}
The enigmatic near-infrared transient VVV-WIT-06 underwent a large amplitude eruption of unclear origin in July 2013. Based on its lightcurve properties and late-time post-outburst spectra various possibilities have been proposed in the literature for the origin of the object viz. a Type I supernova, a classical nova (CN) and a violent stellar merger event. We show that out of these possibilities, an origin in a CN outburst  convincingly explains the observed properties of VVV-WIT-06. We estimate that the absolute $K$ band magnitude of the nova at maximum was M$_{\rm k}$ = -8.2 $\pm$ 0.5, its distance $d$ = 13.35 $\pm$ 2.18 Kpc and the extinction A$_{\rm v}$ = 15.0 $\pm$ 0.55 magnitudes.
\end{abstract}


\section{Introduction} \label{sec:intro}

We analyze the enigmatic Infrared (IR) Transient VVV-WIT-06 discovered by the VISTA Variables in the Via Lactea (VVV) ESO Public Survey \citep{2010NewA...15..433M}. The object showed a large amplitude ($\Delta$K$_{\rm s}$ $>$ 10.5 mag) outburst in 2013 peaking at K$_{\rm s}$ $\sim$ 9 mag during 2013 July and subsequently fading to K$_{\rm s}$ $\sim$ 16.5 in 2017. Analysis of its near-IR spectra obtained  almost 1300d after the eruption by \cite{2017ApJ...849L..23M} showed the spectrum to be heavily reddened and with several emission lines. Based on the light curve and the spectra, \cite{2017ApJ...849L..23M} attempted to classify the object but a definitive classification could not be reached. They proposed three possibilities for its nature viz (i) the closest Type I supernova (SN) observed in about 400 years, (ii) or an exotic high amplitude nova belonging to a regime separate from normal CNe, (iii) a stellar merger event.

Our motivation to reanalyze the object was to resolve its true nature. Furthermore, peculiar objects can sometimes be flag bearers for a new or unique class of objects e.g. as V838 Mon was for stellar mergers \citep{2002A&A...389L..51M} or as V445 Puppis was a prototype for helium novae \citep{2003A&A...409.1007A}. In the present study, we therefore reanalyze some of the \cite{2017ApJ...849L..23M} data and also  present two additional spectra. We arrive at the conclusion that VVV-WIT-06 is a CN whose first spectrum albeit was recorded unusually late, several years after its eruption.

\section{Observations}

Three NIR spectra of VVV-WIT-06  were obtained on 7 March 2017, 26 March 2017 and 21 April 2017 using the Folded Port Infrared Echellette (FIRE) spectrograph on the 6.5m Magellan Baade Telescope \citep{2013PASP..125..270S}. The FIRE spectra were obtained in the high-throughput prism mode with a 0.6 arcsecond slit. The configuration yields a continuous wavelength coverage from 0.8 to 2.5 micron with a resolution of $\sim$ 500 in the J band. At each epoch, an A0V telluric standard was observed close to the science target in time, angular distance and airmass for telluric correction, as per the method described in \cite{2003PASP..115..389V}. Additional details of the observation and reduction steps can be found in \cite{2015A&A...578A...9H}. The log of the observations is given in Table \ref{table-NIRSpec}.  We also show near-IR comparison spectra of the novae V959 Mon (Nova Mon 2012) and V5668 Sgr which were taken at R=1000 (0.85 to 2.4 micron) using the Near-infrared camera and spectrograph (NICS) from the Mount Abu Observatory \citep{2012BASI...40..243B}. Details of observational and data reduction procedures related to NICS are given for e.g. in \cite{2014ApJ...785L..11B,2015MNRAS.452.3696J,2016MNRAS.462.2074S}.

\begin{table}
\centering
\caption{Log of the NIR spectroscopy }\label{table-NIRSpec}
\begin{tabular}{lcccccccc}
\hline
Date & Days & Int. & Std. & Target & Std. \\
 (in 2017) & after max$^a$. & time(s) & star & airmass & airmass \\
\hline
\hline
 March 7 &  1324 & 1014 &  HD 151075 &1.03&1.02\\
 March 26 & 1343 & 3170 &  HIP 75418 &1.03&1.02\\
 April 21 & 1369 & 1390 &  HD 151075 &1.07&1.07 \\
\hline
\hline
\end{tabular}
\begin{list}{}{}
\item (a) Maximum has been taken to occur on 2013 July 22 (MJD 56496).
\end{list}
\end{table}


\section{Results and discussion} 

\begin{figure}
\centering
\includegraphics[width=\columnwidth]{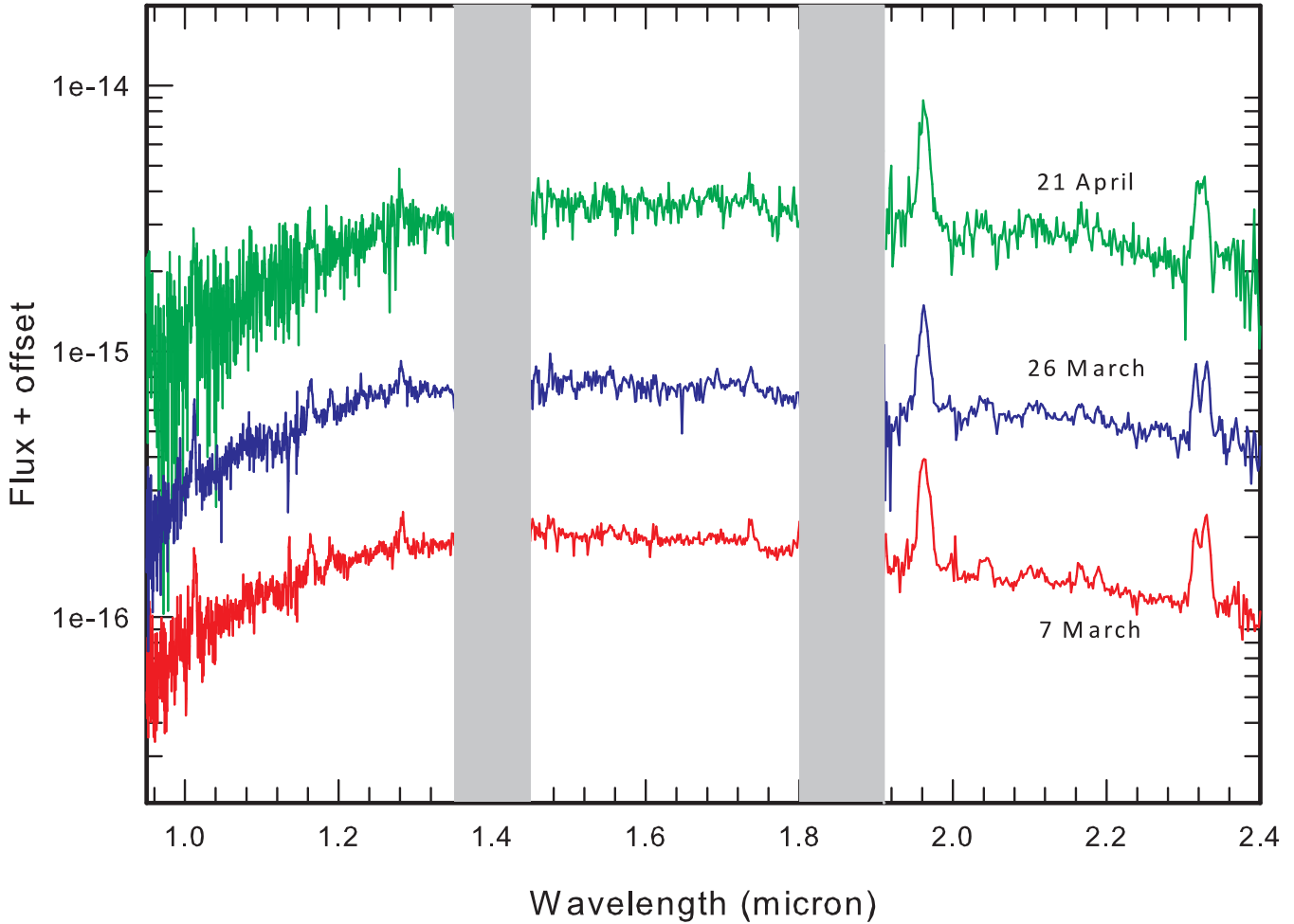}
\caption{Undereddened spectra of VVV-WIT-06 recorded on 7 March 2017, 26 March 2017 and 21 April 2017, respectively.}
\label{fig-MMRD1}
\end{figure}

\begin{figure*}
\centering
\includegraphics[width=\columnwidth]{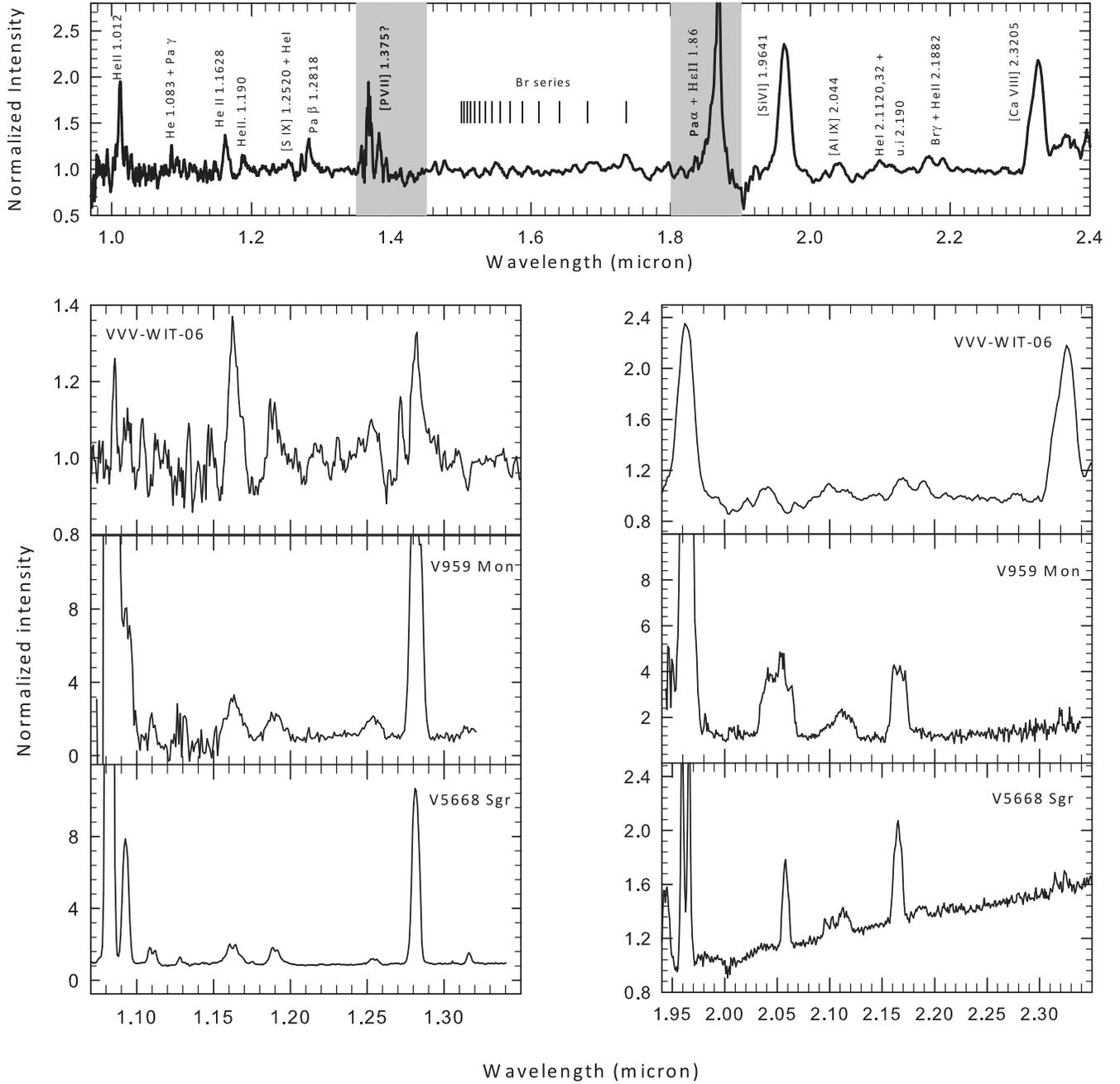}
\caption{The top panel shows the average of the three spectra of VVV-WIT-06 from 7 March 2017, 26 March 2017 and 21 April 2017 after a continuum subtraction and normalization. Regions of poor atmospheric transmission are shown in grey. Lines are marked; the unidentified line at 2.090 $\mu$m blended with He I 2.1120, 2.1132 $\mu$m has sometimes been attributed to [Mn XIV] 2,092 $\mu$m \protect\citep{1996ApJ...467..860W}. Bottom six panels show expanded views of the $J$ and $K$ band spectra of VVV-WIT-06 and also show for comparison the spectra of two other novae with coronal lines viz V959 Mon (He/N type) and V5668 Sgr ( Fe II type; \protect\citealt{2012ATel.4542....1B,2016ATel.8753....1B}). Although the lines in the VVV-WIT-06 spectrum are weak due to the faintness of the object when it was observed, there is a good match between the spectra of all three objects. This strongly suggests that VVV-WIT-06 is a CN. }
\label{fig-MMRD2}
\end{figure*}

Figure \ref{fig-MMRD1} shows the three individual spectra of VVV-WT-06. Since they are reasonably similar (except for the [Ca VIII] 2.3205 $\mu$m line which is discussed later), they were coadded to increase the S/N. The coadded spectrum, after continuum subtraction and normalization, is presented in Figure  \ref{fig-MMRD2} with the lines identified. The spectrum is very typical of a nova in the nebular/coronal stages which is reached when the central white dwarf has evolved to high enough temperatures to emit sufficiently hard radiation to produce ions with high levels of ionization such as Si VI, Ca VIII, Al IX etc (the ionization potentials for these are 166.77, 127.20 and 284.66 eV, respectively). At this stage forbidden transitions dominate the spectrum together with from permitted lines of H and He I and He II. We also show the $J$ and $K$ band spectra of two novae \citep{2012ATel.4542....1B,2016ATel.8753....1B} at a similar stage of ionization. These are V959 Mon which had its outburst in 2012 June 22-24 and V5668 Sgr which erupted on 2015 March 15. The former is an He/N nova \citep{1992AJ....104..725W} while the latter is an Fe II type nova. As can be seen, there is a good match of the spectra and the same lines are seen in the spectra of all three objects. This strongly supports our contention that VVV-WIT-06 is a CN. Any observed difference in the relative strengths of the lines is likely due to the fact that the VVV-WIT-06 observations were made at +1300d after outburst while those for V959 Mon and V5668 Sgr are at $\sim$ +200d and +350d after outburst. The strengths of the recombination lines of H and He are proportional to the square of the electron density and are hence time-dependent and expected to weaken as the ejecta expands and dilutes. The strength of the forbidden lines are also greatly dependent on the density (and hence on the epoch of observation) since they become prominent only below a certain critical density which varies from ion to ion.

\begin{figure}
\centering
\includegraphics[width=0.8\columnwidth]{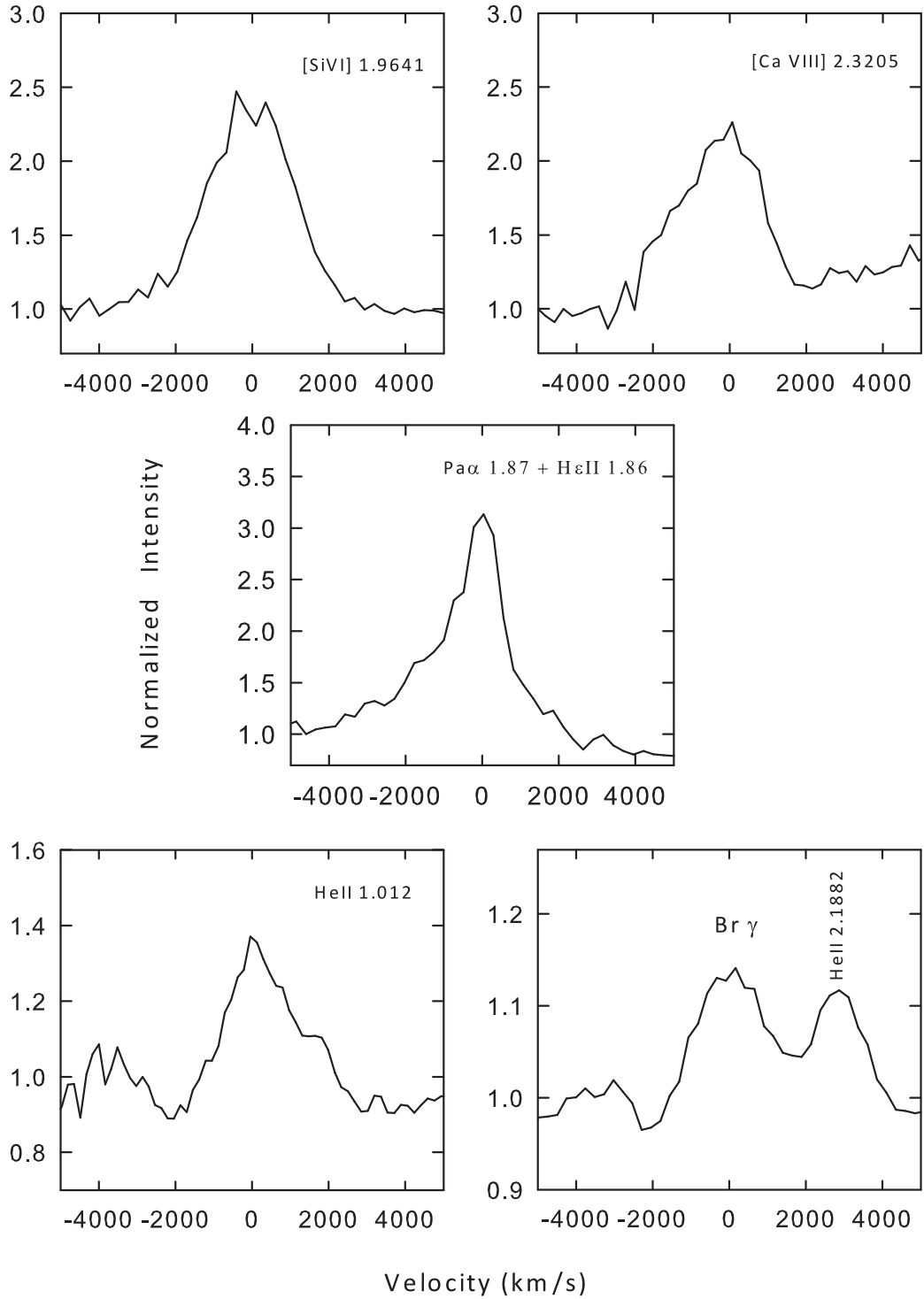}
\caption{Velocity profiles of selected forbidden and permitted lines.}
\label{fig-MMRD3}
\end{figure}

Figure \ref{fig-MMRD3} shows the velocity profiles for a few selected forbidden and permitted lines. The forbidden lines like [Si VI] 1.9641 $\mu$m and [Ca VIII] 2.32 $\mu$m have been recorded at reasonably good S/N to estimate that their full width at zero intensity (FWZI) is $\sim$ 6000 km/s. However all the permitted lines are fairly weak in strength and we had to use a 3 point running average to smooth their profiles in Figure \ref{fig-MMRD3}. Even then the FWZIs of these lines are difficult to ascertain with satisfactory accuracy but it is however clear that they are narrower than the forbidden lines and have a FWZI of $\sim$ 4000-5000 km/s. These values are satisfactorily consistent with generally observed profile velocities in novae. In the classification scheme by \cite{1992AJ....104..725W}, the Fe II novae have FWZI $<$ 5000 km/s whereas the He/N novae have FWZI values generally greater than 5000 km/s. Based on this criterion alone it is however difficult to say whether VVV-WIT-06, if indeed a CN, is of the Fe II type or He/N type. But we would favor the former class because of the extended climb to maximum seen in the light curve and also since the line profiles are not significantly flat-topped as expected for He/N novae.

Another argument suggesting a Fe II class for the object is its position with respect to the galactic plane. Given that the object's galactic latitude is $b$ = 0.88522 degrees and estimated distance $d$ = 13.35 $\pm$ 2.18 kpc, the object's vertical height $z$ above the galactic midplane is 206 $\pm$ 33 parsec. \cite{1998ApJ...506..818D} found that novae belonging to the He/N class tend to concentrate close to the Galactic plane with a typical scale height $\leq$ 100 pc, whereas the Fe II novae are distributed more homogeneously up to $z\leq$ 1000 pc, or even beyond. Although a classification of the object  (He/N or Fe II) is not easy, it may be noted in general that the FWZI values of $\sim$ 5000-6000 km/s observed here, though slightly large,  have been similarly observed in several classical novae (e.g. \citealt{1991IAUC.5223....1H,2011ApJS..197...31S}) thereby supporting our nova  interpretation.

\begin{figure}
\centering
\includegraphics[width=\columnwidth]{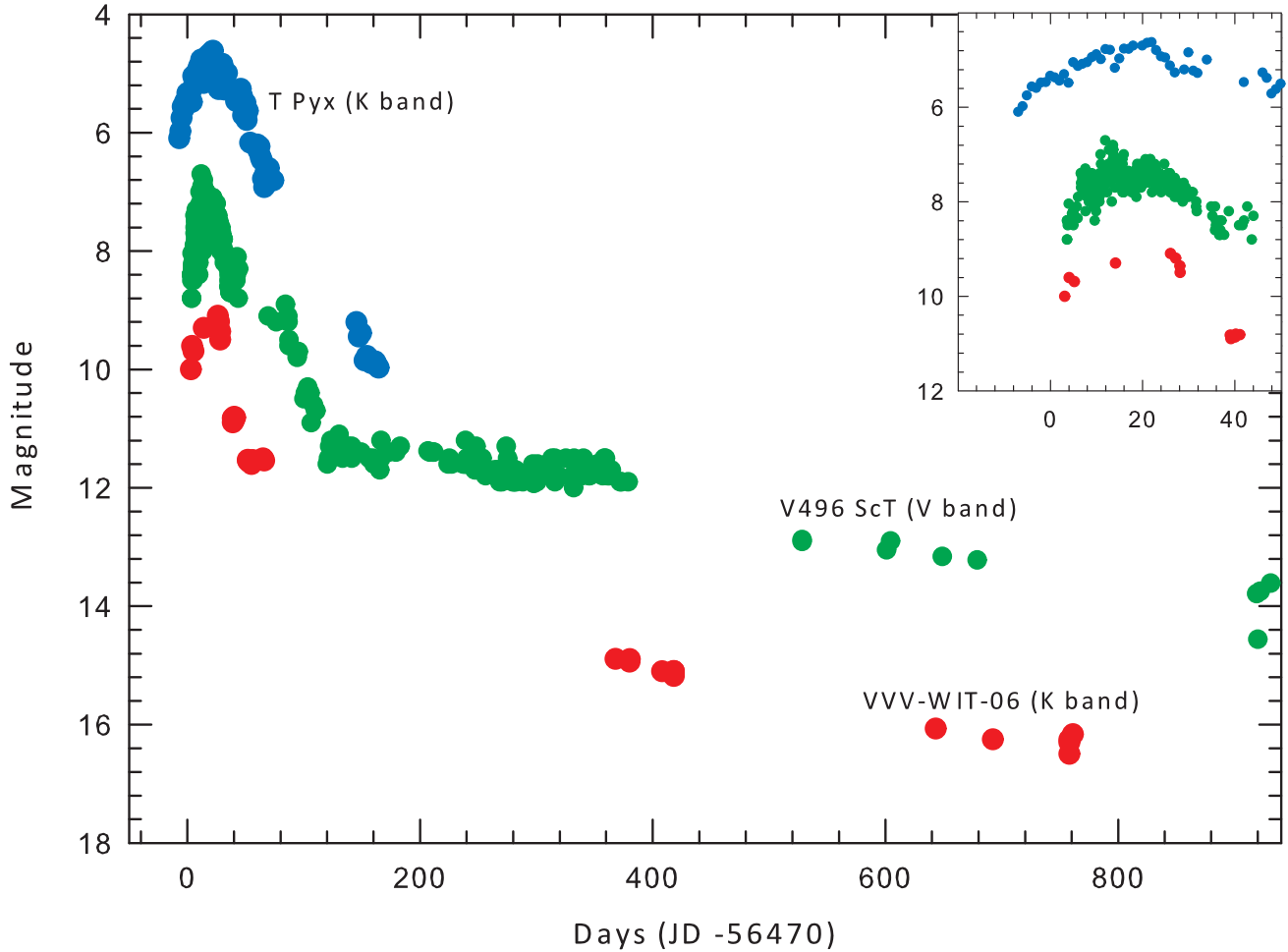}
\caption{The K$_{\rm s}$ band lightcurve of VVV-WIT-06 is shown and for comparison the lightcurves of the novae V496 Scuti (a Fe II type) and T Pyxidis (a recurrent nova) are also shown. The inset, giving an  expanded view of the early rise and decline stages, shows the considerable similarity between the objects.}
\label{fig-LC-RI}
\end{figure}

It is necessary to check whether the characteristics of the light curve are also consistent with the proposed nova classification for VVV-WIT-06 as suggested by the spectra. Figure \ref{fig-LC-RI} shows the K$_{\rm s}$ band light curve from the data available in \cite{2017ATel10140....1M} and for comparison the lightcurves of the novae V496 Scuti and T Pyxidis are also shown. Novae show an enormous variety in their light curve shapes as can be seen from the complilation by \cite{2010AJ....140...34S}. There is also a huge spread in the so called speed class as measured by the t$_{\rm 2}$ parameter, the time for the brightness to decline by two magnitudes from maximum, which ranges from a very few days for the very fast novae to several hundred days for very slow novae. The VVV-WIT-06 lightcurve is a typical nova light curve, with an initial rise to maximum followed by a rapid decline. The slow rise to maximum hints that this is likely a nova of the Fe II class (e.g. V1280 Sco -- \citealt{2008MNRAS.391.1874D}, V339 Del--- \citealt{2017MNRAS.466.4221E,2015ApJ...812..132G}), where such behaviour is often encountered, rather than a He/N nova where the rise and decline from maximum are much more rapid. From the lightcurve we estimate that t$_{\rm 2}$ is 17.5 $\pm$ 0.5 days making this a nova of the fast speed class. This translates to an absolute K$_{\rm s}$ band magnitude M$_{\rm k}$ = -8.2 $\pm$ 0.5 using the maximum magnitude versus rate of decline (MMRD) relation of \cite{1995ApJ...452..704D}. This is very typical value of the absolute magnitude expected for a nova \citep{1995CASSS...5.....H}.

\begin{figure}
\centering
\includegraphics[width=\columnwidth]{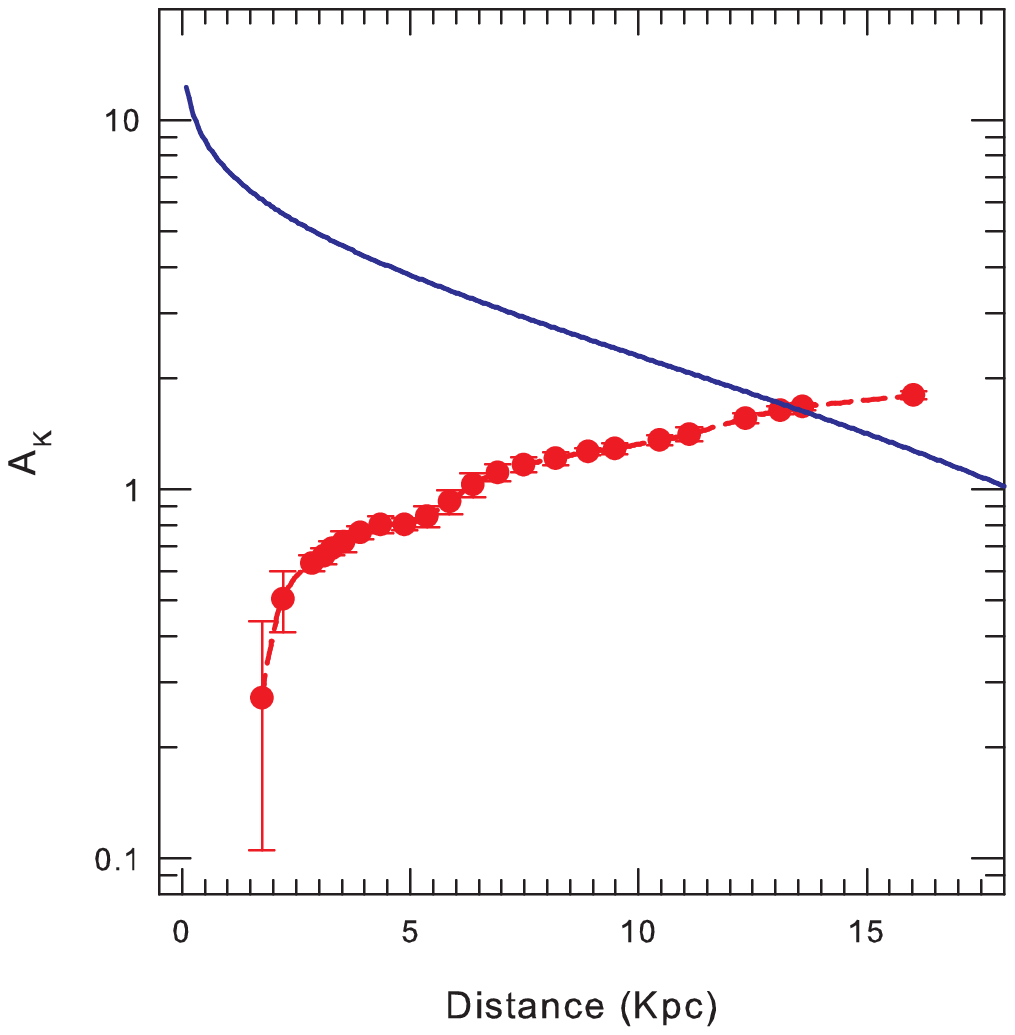}
\caption{The curve joining the data points (red circles) shows the variation of the extinction towards VVV-WIT-06 based on results from \protect\cite{2006A&A...453..635M}. The continuous straight line is a plot of extinction $A_K$ versus distance $d$ from the equation m$_{\rm k}$ - M$_{\rm k}$ = 5 log d - 5 + A$_{\rm k}$ where m$_{\rm k}$ is known from observations and M$_{\rm k}$ is estimated from the MMRD relation. The intersection of the two curves allows to simultaneously estimate the extinction and distance to the nova. More details are given in the text.}
\label{fig-MMRD5}
\end{figure}

We estimate the extinction and distance to the nova in the following way. If a reliable extinction versus distance information is available in the nova's direction (e.g as the \citealt{2006A&A...453..635M} data shown in Figure~\ref{fig-MMRD5}) and a valid MMRD relation for the nova is also available, then estimates of both the extinction and distance can be simultaneously made. The continuous curve in Figure~\ref{fig-MMRD5} shows the A$_{\rm k}$ versus distance curve from the relation m$_{\rm k}$(max) - M$_{\rm k}$ = 5 log d - 5 + A$_{\rm k}$ wherein we use m$_{k}$(max) = 9.09 and M$_{\rm k}$ = -8.2. The second curve in Figure~\ref{fig-MMRD5} shows extinction versus distance from \cite{2006A&A...453..635M} based on their modeling of the galactic extinction. The intersection of the two curves should give the nova's distance and extinction because both curves in principle correspond to the same line of sight and thus the same extinction. We obtain $d$ = 13.35 $\pm$ 2.18 kpc and A$_{\rm k}$ = 1.68 $\pm$ 0.18. Using A$_{\rm k}$/A$_{v}$ = 0.112 and A$_{\rm v}$ = 3.09$E(B-V)$ \citep{1985ApJ...288..618R} we get A$_{\rm v}$ = 15.0 $\pm$ 0.55 and $E(B-V)$ = 4.85 $\pm$ 0.18. These are in reasonably good agreement with similar estimates made by \cite{2011ApJ...737..103S} or \cite{1998ApJ...500..525S} who find $E(B-V)$ values of 4.12 $\pm$ 0.20 and 4.80 $\pm$ 0.23 respectively in the direction of the nova. Some caveats in our calculations includes the fact that the extinction estimates of \cite{2006A&A...453..635M}, \cite{2011ApJ...737..103S} or \cite{1998ApJ...500..525S} are for small regions in which both the extinction and the ratio of selective-to-total extinction could vary (e.g. \citealt{2017ApJ...849L..13A}), and also that near the galactic plane ($|b|$ $<$ 5deg) the extinction value is quite uncertain.

 The split profile of the [Ca VIII] 2.3205 $\mu$m line warrants some discussion. The clear splitting of the profile peak into a blue and a red component is clearly seen in Figure \ref{fig-MMRD1} but unfortunately gets suppressed in averaging spectra in Figure \ref{fig-MMRD2}. None of the other lines show a similar pronounced double peaked structure. The depth and presence of the splitting varies with epoch which may partially be caused by the varying S/N of the three spectra (e.g the April 21 spectrum is noisier and the splitting least pronounced). For understanding the origin of this profile we compare  the [Ca VIII] 2.3205 $\mu$m line profile with other coronal lines. The coronal lines are forbidden transitions that get de-excited by collisions once a critical density is exceeded. For the [Si VI] 1.9641 $\mu$m line the critical density is 10$^{\rm 8}$ cm$^{\rm -3}$ (see \citealt{2014IAUS..304..180S,1992ApJ...399..504S}). The critical density for the [Ca VIII] 2.3205 $\mu$m line is not covered by this work. However, \cite{1997ApJS..110..287F} analysing coronal line strengths in AGN's show that the [Ca VIII] 2.3205 $\mu$m line is not present for densities exceeding 10$^{\rm 6.25}$ cm$^{\rm -3}$ while the [Si VI] 1.9641 $\mu$m line is not seen beyond a density of 10$^{\rm 8}$ cm$^{\rm -3}$. We thus assume the critical density of the [Ca VIII] 2.3205 $\mu$m line is 10$^{\rm 6.25}$ cm$^{\rm -3}$. Hence it should arise from regions of considerably lower densities compared to the [Si VI] 1.9641 $\mu$m line i.e from different sites within the nova which could therefore have different kinematic properties. One possibility is that VVV-WT-06 has a bipolar morphology - many novae have such a shape as discussed in great detail in \cite{2018MNRAS.473.1895B}. In such a geometry the material in the equatorial plane is dense and slow moving while the density in the polar regions is low since the polar regions expand fast and dilute quickly. Thus the Si VI and other atomic emission could emanate from all over the nova including the equatorial waist whereas the [Ca VIII] 2.3205 $\mu$m would preferably be located mostly in the polar regions. The blue and red components would then correspond to the line of sight velocity components from the approaching and receding polar regions of the bipolar structure. An alternative possibility for a dip in the central region of the peak is that it is caused by self-absorption since the line is fairly strong.

\section{Discussion}

If the observed light curve represents that of a CN, the derived quantities from it such as t$_{\rm 2}$ and the absolute magnitude are also consistent with those expected for a CN. The estimated extinction, matches the observed extinction in the direction of the nova. The amplitude of outburst, from quiescence to peak brightness, of $>$ 10.5 mags is consistent with what is expected for a nova; the slowest novae have an amplitude around 7 while the fastest around 15 \citep{1995CASSS...5.....H,2008clno.book.....B}. The distance estimate also appears reasonable being within the Milky Way. Thus, the evidence supports a CN classification. The spectra independently also strongly support the CN nature.

\cite{2017ApJ...849L..23M} have considered two alternative eruption scenarios which could produce such a high amplitude outburst as observed in VVV-WIT-06. One of the possibilities is that it is a Type Ia SN. 
But as the authors admit, there are several difficulties in assigning an SN origin to VVV-WIT-06. These include the fact that the late evolution of the lightcurve departs from the characteristic power-law radioactive decay of SN Ia because the brightness of VVV-WIT-06 remains nearly constant during 2016 and 2017, indicating an additional energy source. Additionally NIR spectra of type Ia SNe are not available up to $\sim$ 1300d after explosion not allowing a direct comparison with the observed late spectra of VVV-WIT-06. The other problems with a SN interpretation are the low probability ($<$ 1 percent) that the host galaxy would align within less than a degree with the Galactic plane and that the measured kinematics (radial velocity and proper motions) of VVV-WIT-06 favor a Galactic origin.

The other possibility considered is that it could be a stellar-merger event. The observational realm of stellar mergers took birth in 2002 with the spectacular outburst of haloed V838 Mon. 
It was soon realized that two other objects, V4332 Sgr and M31-RV in Andromeda, shared similar evolutionary properties with V838 Mon leading to the proposition that they comprised a new class of objects \citep{2002A&A...389L..51M,2002A&A...395..161B}. 
However all these objects showed a common trend of evolving towards cool temperatures after their outburst (e.g. \citealt{2004ApJ...607..460L,2002A&A...395..161B,2003MNRAS.343.1054E,1989ApJ...341L..51R}). Their post-outburst spectra resembled that of cool stars replete and rich with molecular bands, either in absorption or emission, of many species like CO, AlO, H$_{\rm 2}$O, TiO, ScO, CrO etc \citep{2004ApJ...607..460L,2003ApJ...598L..31B,2004ApJ...604L..57B,2004ApJ...615L..53B,2005ApJ...627L.141B,2015A&A...580A..34K}. Many of them formed copious dust that exists till date \citep{2015ApJ...814..109B,2014A&A...569L...3C}. This is in complete contrast with VVV-WIT-06 whose ejecta has evolved in the opposite direction viz towards high ionization stages containing species like Si VI, Al IX etc. We thus consider the stellar merger scenario to be completely ruled out and in view of earlier arguments propose a CN origin for VVV-WT-06.

\acknowledgments

We thank an anonymous reviewer  for helpful comments. We thank Dr. Mark Phillips for sparing some of his telescope time for these observations and also commenting on the manuscript. E.~Y.~H. acknowledges the support provided by the National Science Foundation under Grant No. AST-1613472 and by the Florida Space Grant Consortium. LG acknowledges support from the FINCA visitor programme. The research work at the Physical Research Laboratory is funded by the Department of Space, Government of India.

\facility{Magellan:Baade(FIRE).}

\bibliographystyle{aasjournal}
\bibliography{vvv}

\end{document}